\documentclass[sigconf]{acmart}

\AtBeginDocument{%
  }

\begin{document}

\acmYear{2026}\copyrightyear{2026}
\setcopyright{cc}
\setcctype[4.0]{by}
\acmConference[ASIA CCS '26]{ACM Asia Conference on Computer and Communications Security}{June 1--5, 2026}{Bangalore, India}
\acmBooktitle{ACM Asia Conference on Computer and Communications Security (ASIA CCS '26), June 1--5, 2026, Bangalore, India}
\acmDOI{10.1145/3779208.3804894}
\acmISBN{979-8-4007-2356-8/26/06}

\title{Phantom Force: Injecting Adversarial Tactile Perceptions into Embodied Intelligence via EMI}

\author{Zirui Kong}
\orcid{0009-0001-8918-7822}
\affiliation{%
  \institution{The Hong Kong Polytechnic University}
  \city{Hong Kong}
  \country{China}}
\email{jerryhongkong.kong@connect.polyu.hk}

\author{Youqian Zhang}
\orcid{0000-0003-0907-7998}
\affiliation{%
  \institution{The Hong Kong Polytechnic University}
  \city{Hong Kong}
  \country{China}}
\email{you-qian.zhang@polyu.edu.hk}

\author{Sze Yiu Chau}
\orcid{0000-0001-9300-0808}
\affiliation{%
  \institution{Simon Fraser University}
  \city{Burnaby}
  \country{Canada}}
\email{sychau@sfu.ca}

\begin{abstract}
Embodied intelligent robots rely on tactile sensors to interact with the physical world safely. 
While the security of visual perception systems has been studied (e.g., adversarial samples), the integrity of the tactile sensory channel remains unexplored. 
This work explores a vulnerability in Hall-effect fingertip sensors, showing their susceptibility to intentional Electromagnetic Interference (EMI). 
We demonstrate that a targeted signal injection can induce strong ``phantom forces'', amplifying perceived force magnitude by over \textbf{9$\times$} and deviating the inferred force direction by \textbf{65$^\circ$}.
Such perturbations can paralyze learning-based tactile classification models, seriously affecting robot movement. 
An attacker could exploit this vulnerability to coerce a robot hand into crushing fragile objects or dropping dangerous payloads.
\end{abstract}



\settopmatter{printfolios=false}

\maketitle

\section{Introduction}

\begin{figure}[t]
\centering
\includegraphics[width=0.5\columnwidth]{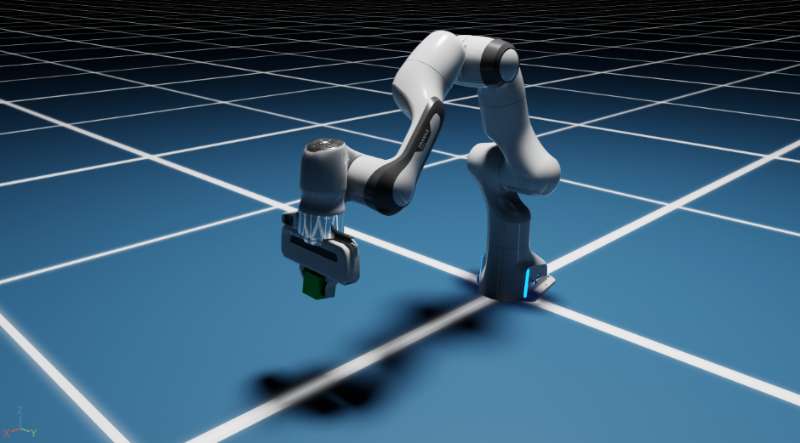}
\caption{Simple attack experiment simulation in NVIDIA Isaac Sim \cite{NVIDIA_Isaac_Sim}. The Franka Emika Panda robot is equipped with a Hall-effect based tactile sensor.}
\label{fig:setup}
\end{figure}

Embodied intelligence refers to intelligent behavior emerging not solely from abstract computation, but from the tight, real-time coupling between an agent’s physical body and the physical world~\cite{10.7551/mitpress/3585.001.0001}. 
In the last few years, this paradigm has gained substantial traction in both academic and industrial robotics, enabling significant advances in dexterous manipulation, adaptive control, and human-robot interaction~\cite{10415857}. 
In these contact-rich environments, the robotic hand (e.g., Fig.~\ref{fig:setup}) serves as a primary interface for interaction, relying heavily on fingertip tactile sensors to perceive object properties such as contact force, pressure distribution, and material texture.

However, while much attention has been paid to improving the functionality and precision of these tactile sensors, their security has been largely ignored. 
In most current systems, signals from tactile sensors are treated as inherently trustworthy. 
This assumption of signal integrity creates a critical vulnerability: an attacker could inject malicious signals at the sensor level.
In this work, we focus on the security of modern tactile sensors, specifically finger-mounted Hall-effect sensors. 
We demonstrate the following findings:
(1) \textbf{Adversarial Perturbation of Force Estimates:} Carefully crafted electromagnetic waves can manipulate the tactile sensor's force readings, altering both the magnitude and direction of the estimated contact force.
(2) \textbf{Impact on Machine Learning-Based Task:} Such perturbations propagate through learned models, leading to adverse performance degradation.

\section{Methodology}

We first establish a tactile classification baseline using controlled weights, and then quantify the impact of malicious electromagnetic interference (EMI) on the same task.

\subsection{Data Acquisition and Baseline Task}

\begin{figure}[t]
\centering
\includegraphics[width=0.9\columnwidth]{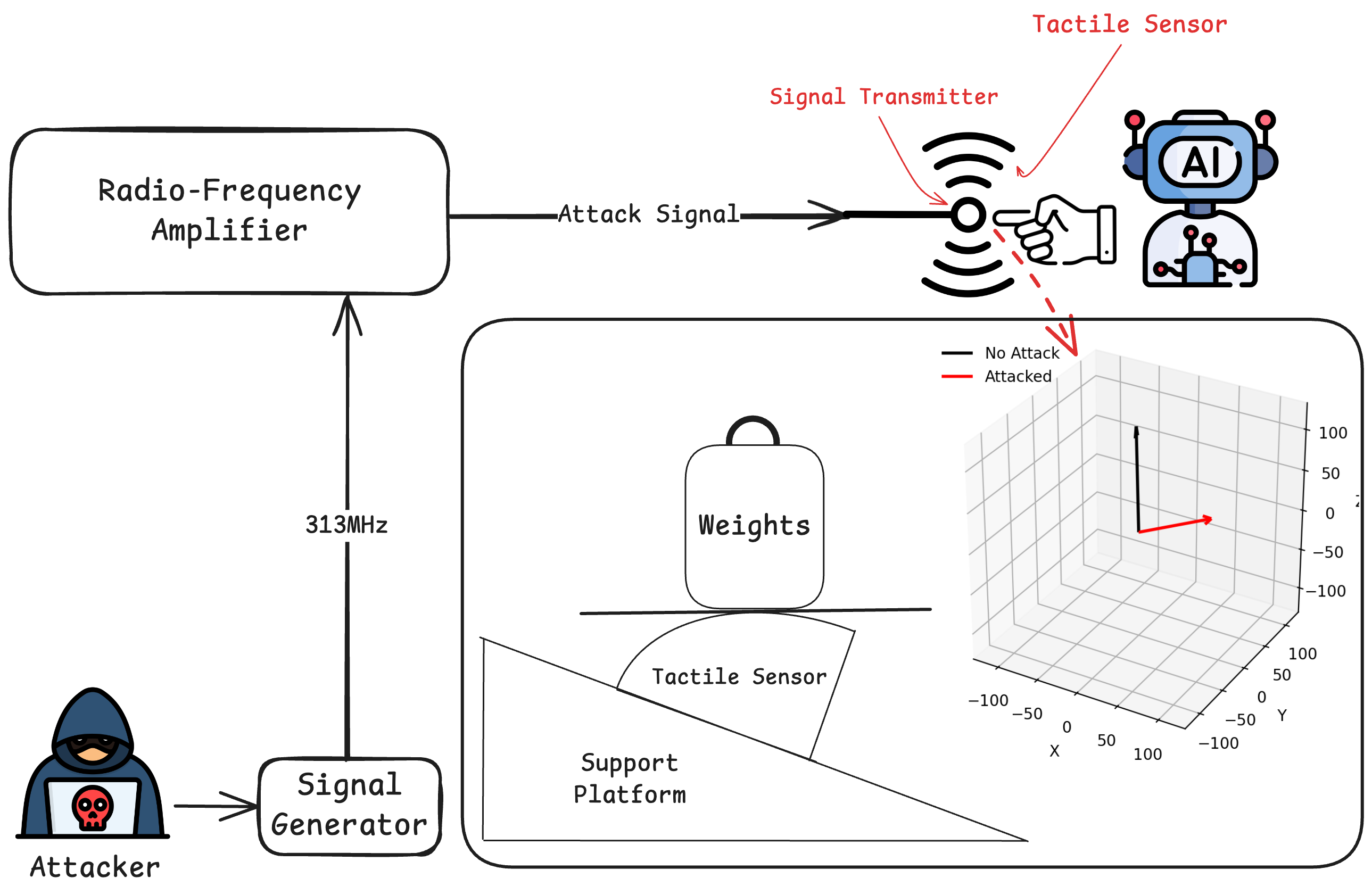}
\caption{Attack experiment model and setup in reality.}
\label{fig:model}
\end{figure}

We emulate fingertip tactile inputs by pressing discrete calibration weights onto the Gen. 2 Paxini Hall-effect fingertip sensor mounted on the robot end-effector. 
The experimental setup is visualized in Fig~\ref{fig:model}. For each weight, we collect time-series sensor readings at a fixed sampling rate under a standardized contact protocol. 
Three-dimensional raw signals are synchronized and segmented into windows to train a Random Forest (RF) classifier.

\subsection{Attack Setup and Evaluation}
To probe vulnerability at the physical layer, as shown in Fig.~\ref{fig:model}, we deploy a high-frequency emitter centered at 313 MHz, identified through a prior 100–400 MHz sweep as a resonant frequency that induces measurable perturbations on the fingertip sensor signal. 
The attack chain consists of a signal generator producing a continuous-wave tone, followed by a power amplifier, and a near-field probe positioned at a controlled standoff distance ($<$1 cm) to the sensor. 
Output power at the probe tip is calibrated using a spectrum analyzer and a reference probe. 

We evaluate classification degradation relative to the baseline.
At the same time, we evaluate two key metrics: 

\paragraph{Cosine Similarity (Cos. Sim.)}

The cosine similarity measures the angular alignment between the measured force vector and the ground-truth force vector. It is defined as:
\(
\text{Cos. Sim.} = 
\frac{
    \mathbf{F}_{\text{gt}} \cdot \mathbf{F}_{\text{measure}}
}{
    \left\| \mathbf{F}_{\text{gt}} \right\| 
    \cdot 
    \left\| \mathbf{F}_{\text{measure}} \right\|
}
\),
where \( \mathbf{F}_{\text{gt}} \) is the ground-truth force vector, and \( \mathbf{F}_{\text{measure}} \) is the measured (possibly perturbed) force vector. The result ranges from –1 (opposite direction) to 1 (same direction), with 0 indicating orthogonality.

\paragraph{Amplitude Ratio (Amp. Rat.)}

The amplitude ratio quantifies the difference in magnitude between the measured and ground-truth force vectors:
\(
\text{Amp. Rat.} = 
\frac{
    \left\| \mathbf{F}_{\text{measure}} \right\|
}{
    \left\| \mathbf{F}_{\text{gt}} \right\|
}
\).
A ratio greater than 1 indicates amplification of the force magnitude due to perturbation, while a ratio less than 1 indicates attenuation.

\section{Experimental Results and Discussion}

As shown in Table~\ref{tab:tactile_attack}, when no attack signal is present, the force measured by the tactile sensor closely matches the ground truth, where the Cos. Sim. and Amp. Rat are above 0.99, and where precision (P), recall (R), and F1 score (F1) are around 0.95, indicating fairly accurate detection of the force. However, when an attack occurs, it can simultaneously alter both the direction and magnitude of the measured force.
Such perturbations for RF classifier lead to a complete collapse in performance, with both precision and recall dropping to zero, as shown in Table~\ref{tab:tactile_attack}. The drastic change in signal amplitude moves the data distribution entirely outside the learned boundaries of the model.

\begin{figure}[t]
\centering
\includegraphics[width=0.99\columnwidth]{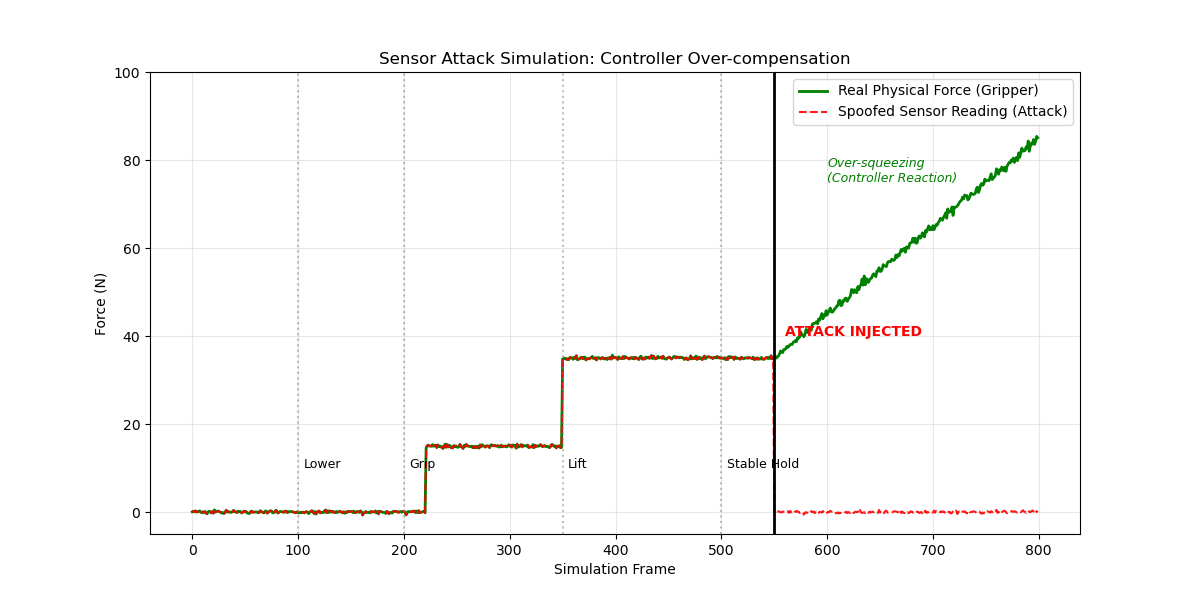}
\caption{Sensor Attack Simulation during the Stable Hover Phase. The green line represents the real physical force exerted by the gripper (approx. 35N). The red dashed line shows the spoofed sensor reading dropping to near 0N at Frame 550 (``Attack Injected''), simulating a false ``loss of grip'' event.}
\label{fig:chart}
\end{figure}

Further, we conducted a simulated time-domain analysis of the attack on NVIDIA Isaac Sim platform, as shown Fig.~\ref{fig:chart}. 
During the ``Stable Hold'' phase (Frames 350-550), the robot maintains a steady grip with a real force of approximately 35N. 
At Frame 550, the attack is injected. While the \textit{Real Physical Force} (green line) remains constant, the \textit{Spoofed Sensor Reading} (red dashed line) drastically drops to zero. 
This type of attack creates a ``Phantom Release'' scenario: the robot believes it is about to drop the object, while in reality it is still holding it. In a closed-loop control system, this would typically trigger the robot to increase its gripping force dangerously to ``regain'' the object, potentially crushing it.

\begin{table}[t]
\centering
\caption{Performance metrics under benign and adversarial cases. Mean ($\mu$) and standard deviation ($\sigma$) are reported.}
\label{tab:tactile_attack}
\begin{tabular}{llllll}
\toprule 
\textbf{Cases} & \textbf{Cos. Sim.} & \textbf{Amp. Rat.} & \textbf{P} & \textbf{R} & \textbf{F1} \\
& $\mu$ ($\sigma$) & $\mu$ ($\sigma$) & $\mu$ ($\sigma$) & $\mu$ ($\sigma$) & $\mu$ ($\sigma$) \\
\midrule 

Non-Attack 
    & 0.99 & 1.00 & 0.95 & 0.95 & 0.95 \\
    & (0.0003) & (0.0003) & (0.05) & (0.08) & (0.04) \\

\addlinespace[1em] 

Attack 
    & 0.56 & 9.20 & 0.00 & 0.00 & -- \\
    & (0.44) & (8.90) & (0.00) & (0.00) & \\
\bottomrule 
\end{tabular}
\end{table}

\section{Conclusion and Future Work}

In this preliminary study, we empirically demonstrated that, through real-world experiments and Isaac Sim simulations, fingertip tactile sensors in embodied intelligent systems are vulnerable to EMI attacks.
These results expose a previously underexplored attack surface in robotic perception and underscore a critical insight: ensuring the robustness of embodied intelligence requires a holistic view of sensor security, extending beyond vision to include tactile and other proprioceptive modalities. 

Building on these findings, several important scientific directions warrant further investigation. 
First, a systematic characterization of adversarial signal injection is needed to understand its fundamental capabilities and limitations, including attack range, power requirements, temporal characteristics, alternative injection modalities, etc. 
Extending this analysis across diverse tactile sensing technologies, such as capacitive, piezoresistive, optical, and multimodal sensors, will be essential to assess the generality of the threat and to identify modality‑specific weaknesses.
Second, future work should explore principled defense mechanisms that operate across the sensing and learning stack. Promising directions include hardware‑level sensor design, signal‑level anomaly detection and filtering, and learning‑based approaches that improve robustness through sensor fusion, uncertainty modeling, and adversarially informed training. 
Ultimately, these efforts can contribute to a unified framework for trustworthy tactile perception, enabling embodied robots to operate in a secure and safe manner in adversarial noisy environments.

\begin{acks}
This work was partially supported by a grant (Grant Number: P0059492) from The Hong Kong Polytechnic University, and a grant (Grant Number: FRG 272615) from Simon Fraser University.
\end{acks}

\bibliographystyle{ACM-Reference-Format}
\bibliography{references}

@book{10.7551/mitpress/3585.001.0001,
    author = {Pfeifer, Rolf and Bongard, Josh},
    title = {How the Body Shapes the Way We Think: A New View of Intelligence},
    publisher = {The MIT Press},
    year = {2006},
    month = {10},
    isbn = {9780262281553},
    doi = {10.7551/mitpress/3585.001.0001},
    url = {https://doi.org/10.7551/mitpress/3585.001.0001},
}

@ARTICLE{10415857,
  author={Tong, Yuchuang and Liu, Haotian and Zhang, Zhengtao},
  journal={IEEE/CAA Journal of Automatica Sinica}, 
  title={Advancements in Humanoid Robots: A Comprehensive Review and Future Prospects}, 
  year={2024},
  volume={11},
  number={2},
  pages={301-328},
  doi={10.1109/JAS.2023.124140},
  ISSN={2329-9274},
  month={February},}

@software{NVIDIA_Isaac_Sim,
    author = {{NVIDIA}},
    license = {Apache-2.0},
    title = {{Isaac Sim}},
    url = {https://github.com/isaac-sim/IsaacSim},
    version = {5.1.0}
}

\end{document}